\documentclass[usenatbib]{mnras}
\usepackage[pdftex]{graphicx}
\usepackage{latexsym}
\usepackage{color}
\title[The oldest and most metal poor stars in the Galaxy]{The oldest and most metal poor stars in the APOSTLE Local Group simulations} 
\author[E. Starkenburg et al.,]{Else Starkenburg$^1$\thanks{E-mail: estarkenburg@aip.de}
Kyle A. Oman$^2$,
Julio F. Navarro$^2$,
Robert A. Crain$^3$,
\newauthor{Azadeh Fattahi$^2$,
Carlos S. Frenk$^4$,
Till Sawala$^5$,
Joop Schaye$^6$}\\
$^1$ Leibniz Institute for Astrophysics Potsdam (AIP), An der Sternwarte 16,
  14482 Potsdam, Germany\\
$^2$ Dept. of Physics and Astronomy, University of Victoria, P.O. Box 3055, STN CSC, Victoria BC V8W 3P6, Canada\\
$^3$ Astrophysics Research Institute, Liverpool John Moores University, IC2, Liverpool Science Park, 146 Brownlow Hill, Liverpool L3 5RF, U.K.\\
$^4$ Institute for Computational Cosmology, University of Durham, U.K.\\
$^5$ Department of Physics, University of Helsinki, Gustaf H\"{a}llstr\"{o}min katu 2a, FI-00560 Helsinki, Finland\\
$^6$ Leiden Observatory, Leiden University, PO Box 9513, NL-2300 RA Leiden, the Netherlands}
\begin{document}

\maketitle

\begin{abstract}
We examine the spatial distribution of the oldest and most metal poor stellar populations of Milky Way-sized galaxies using the APOSTLE cosmological hydrodynamical simulations of the Local Group. In agreement with earlier work, we find strong radial gradients in the fraction of the oldest (t$_{\rm form} < 0.8$ Gyr) and most metal poor ([{\rm Fe}/{\rm H}]$ < -2.5$) stars, both of which increase outwards. The most metal poor stars form over an extended period of time; half of them form after $z=5.3$, and the last 10\% after $z=2.8$. The age of the metal poor stellar population also shows significant variation with environment; a high fraction of them are old in the galaxy's central regions and an even higher fraction in some individual dwarf galaxies, with substantial scatter from dwarf to dwarf. Overall, over half of the stars that belong to both the oldest and most metal poor population are found outside the solar circle. Somewhat counter-intuitively, we find that dwarf galaxies with a large fraction of metal poor stars that are very old are systems where metal-poor stars are relatively rare, but where a substantial old population is present. Our results provide guidance for interpreting the results of surveys designed to hunt for the earliest and most pristine stellar component of our Milky Way.
\end{abstract}


\section{Introduction}
It is often assumed that the most metal poor
stars today are the oldest objects in the Galaxy -- and as such they are the most extensively
studied relics from the ancient Universe. While very useful as a first
order approximation, it is nevertheless obvious that
the adoption of metallicity as a clock is only valid in closed-box
environments, where metals are well-mixed at all times. In reality, it is
expected that chemical evolution starts and progresses
at different rates in various
environments. To complicate matters even more, the $\Lambda$CDM cosmological paradigm of hierarchical
formation allows stars to
find themselves in very different environments throughout their
lives -- albeit still bearing the chemical imprint of their birth location.  

The very first stars in the
Universe are expected to form in environments that are overdense at early times. In low-density
environments on the other hand, pristine stars may
have formed much later -- provided that the gas in them has remained unpolluted. Simulations suggest that pristine stars can still form at moderate redshifts well after the completion of global reionization \citep[e.g.,][]{Scannapieco03,Scannapieco06,Tornatore07,Trenti09, Muratov13a,Xu16}. A distinction is sometimes made between those stars that were formed out of unpolluted gas but have been affected by radiation or kinematical feedback from other nearby stars, and those that are truly the first stars in their environment \citep[e.g.,][]{McKee08}. 

From a modelling perspective, simulations of the formation of the very first stars have proved to be complex. Generally, it is thought that their initial mass function is more top heavy than that of present day star formation, due to the inability of the initial gas cloud to cool and collapse via metal line emission. 
However, it is unclear at what level later fragmentation of the protostellar cloud could
facilitate the formation of subsolar mass stars which -- even if formed at the
dawn of time -- would still be around today  \citep[see][and references therein]{Karlsson13, Bromm13,Greif15}. Furthermore, a full treatment of
metal-line, molecular, and dust cooling, stellar nucleosynthesis of zero-metallicity stars, their stellar feedback and the
reionization epoch would be needed to
model accurately the epoch of the first stars. None of these processes is
well understood and even when certain assumptions are made, it is 
particularly challenging and computationally expensive to simulate them within a
fully cosmological setting. Modelling studies have therefore predominantly either
focussed on a detailed treatment of processes in the early universe in a
limited volume and/or a limited time \citep[e.g.,][]{Abel02,Wise12,Muratov13a, Muratov13b,Smith15}, or resorted to a simpler
treatment of physical processes - often through a so-called semi-analytical
approach whereby physical processes are modelled by analytical relations on
top of a cosmological merger tree or Press-Schechter formalism \citep[e.g.,][]{Scannapieco03,Yoshida04,Scannapieco06, Salvadori07,Salvadori10, Komiya10,Tumlinson10,Gao10}.  

Observational endeavours have been extremely helpful in constraining
models of early epoch star formation in the Milky Way. Several survey efforts
have been mining the Milky Way stars for the lowest-metallicity
members \citep[e.g.,][]{Beers85, Christlieb03, Keller07, Schlaufman14, Casey15, Howes16} and in the coming decade it is expected that the search for metal poor stars will
intensify. At present, nine stars are known to have $[{\rm Fe}/{\rm H}]<-4.5$ \citep{Christlieb02, Frebel05,
  Norris07, Caffau11, Hansen14, Keller14, Bonifacio15, Frebel15}. With such low numbers of stars at the lowest metallicities, any history of the Milky Way's earliest epochs that might be derived from them would
remain anecdotal at best. Another limitation of observational work in this
field is that age determinations of individual stars have large
uncertainties, up to several gigayears for old ages. 

The best avenue available to
link these stars to the very early Universe is by comparing their abundance patterns to the predicted yields for
  metal-free massive stars. For instance, the most iron-poor star, SMSS J031300.36-670839.3 at $[{\rm Fe}/{\rm H}]$
  $<$ -7.0 could be linked to a $\sim$60 solar-mass, zero-metallicity progenitor
  \citep{Keller14}. It is particularly intriguing that, from analysis of the
  limited numbers of extremely metal poor stars ($[{\rm Fe}/{\rm H}]<-3$), it seems that
  their abundance patterns vary with environment in the Galaxy. Recently,
  \citet{Howes15} showed that in 23 stars in the bulge region
  none shows significant carbon enhancement, whereas this occurs in 
  $\sim$32 per cent of halo stars of such low metallicities
  \citep{Yong13b}. Although always based on small samples, in earlier work it has already been noticed that the number of carbon-enhanced stars (or their abundance patterns) seems to change as a function of environment. Variations -- or hints thereof -- have been reported with increasing disk height \citep{Frebel06},  in the inner versus outer
halo \citep{Carollo14} and in the Sculptor dwarf spheroidal galaxy compared to the halo \citep{Starkenburg13, Skuladottir15}. If confirmed by larger studies, this might indicate that the most metal poor stars in various Milky Way environments have been formed through different channels, or even at different times \citep[see also the discussion in][]{Gilmore13,Norris13,Starkenburg14,Salvadori15}. 

Similarly, the diversity in abundance patterns found among the metal poor stars in satellite galaxies can give us also clues on the early Universe and chemical enrichment processes \citep[one of the more striking examples are the r-process enhanced stars in Reticulum II, ][]{Ji16}. Observationally, dwarf galaxies -- and particularly those as close to us as the satellites of the Milky Way -- are
of great value, because they allow us to study in detail the early formation of galaxies in a different mass scale. Some of the faintest system, belonging to the class of so-called ``ultra-faint'' galaxies, seem to qualify to be ``first galaxies'' or ``fossil galaxies'' -- meaning that the great majority of their total stellar population was formed before the epoch of reionization \citep{Bovill09,Bovill11a, Bovill11b}. However, which and how many systems exactly qualify fossil galaxies criteria is debated and it is unclear to what extent quenching through tidal disturbance by the Milky Way could influence these results \citep{Weisz14}. \citet{Brown14} derives ages from Hubble Space Telescope colour-magnitude diagrams for six ultra-faint galaxies and finds all of them are consistent with having formed 80 per cent of their
stars by $z=6$. Additional evidence for only one
short burst of star formation can come from chemical element analysis of
their stars showing no enrichment by, for instance, supernovae Type\ Ia products
or s-process elements \citep{Frebel12,Frebel14}. 

In this work, we study the early generations of stars
in the APOSTLE simulations \citep{Sawala16,Fattahi16}. This is a suite of twelve volumes selected to resemble the Local Group simulated with sufficient mass resolution
to study the Milky Way and Andromeda-type galaxies as well as their smaller companions and isolated dwarf systems. Moreover, the
hydrodynamical nature of the simulations allows us to follow enrichment and
feedback processes more self-consistently than in semi-analytic
frameworks. The main purpose
of this work is to present further predictions and guidelines for the interpretation of
present and future surveys of metal-poor stars throughout the Galaxy. We will introduce our suite of simulations in more detail in Section \ref{sec:apostle}. We first review the general distribution of metallicity and age within the Milky Way and M31 analogues in Section \ref{sec:metageapostle}, before focussing our attention on the oldest and most metal poor stars in Sections \ref{sec:oldormetpoor} and \ref{sec:oldandmetpoor} where we study their distribution in the main galaxies. Additionally, we compare these findings with the old and most metal poor populations in the surviving satellites and isolated dwarf galaxies in Section \ref{sec:dwarfs}. To aid future surveys, we discuss how to find the systems in which a metal-poor star is most likely to be very old in Section \ref{sec:howto}. We finally discuss the assumptions and a comparison with literature results  in Section \ref{sec:disc}. Throughout this paper we adopt the solar abundance for iron to be ${\rm log} \ \epsilon_{\rm Fe}$ = 7.50\footnote{We use here the customary astronomical scale for logarithmic abundances where log $\epsilon_{\rm X}$ = ${\rm log}_{10}$(N$_{\rm X}$/N$_{\rm H}$)+12, with N$_{\rm X}$ and N$_{\rm H}$ the number
densities of element X and hydrogen, respectively.} \citep{Asplund09}.

\section{The APOSTLE simulations}\label{sec:apostle}

The APOSTLE simulations are described in detail by \citet{Sawala16,
  Fattahi16}. In brief, the
APOSTLE set consist of 12 halo pairs that are selected from
the larger cosmological DOVE simulation volume \citep{Jenkins13} and resemble the Local
Group pair of galaxies in several properties: (i) their spatial separation
(between 600 and 1000~kpc), (ii)
their relative radial velocity (in the range -250 to 0~km~s$^{-1}$) (iii) their relative
tangential velocity (less than 100~km~s$^{-1}$),
(iv) their total pair mass ($\log_{10}(M_{\rm tot}/{\rm M_{\odot}}$)
is between 12.2 and 12.6\footnote{When quoting halo masses we use
  $M_{200}$, the mass enclosed within a spherical volume where the
  mean overdensity is 200 times the critical density.}), and (v) they
are sufficiently isolated, with no haloes larger than the smallest of
the pair within 2.5~Mpc of the midpoint of the pair and a relatively
unperturbed Hubble flow out to 4 Mpc. Interestingly,
\citet{Fattahi16} find that halo pairs that satisfy these criteria are on
average a factor of 2 lighter than the total mass of the Local Group estimated from the timing argument. The median masses of the primary and secondary halo in
each region are $1.4 \times 10^{12}$ M$_\odot$ and $0.9 \times 10^{12}$ M$_\odot$
respectively. Each region samples a high-resolution sphere out to at least 2.5~Mpc, so many more nearby isolated systems within the Local Volume can also be studied. The APOSTLE simulations are run at three resolution levels
labelled ``L1'', ``L2'' and ``L3'' and have approximate dark matter particle masses of 5.0 $\times 10^{4}$ M$_{\odot}$,
5.9 $\times 10^{5}$ M$_{\odot}$, and 7.5 $\times 10^{6}$ M$_{\odot}$ and
approximate primordial gas particle masses of 1.0 $\times 10^{4}$ M$_{\odot}$, 1.2 $\times
10^{5}$ M$_{\odot}$, and 1.5 $\times 10^{6}$ M$_{\odot}$ respectively. Only AP-1, AP-4 and AP-11 have been simulated at the highest (L1) resolution. The simulations at resolutions L2 and L3 have been completed for all twelve volumes. The different resolutions use identical subgrid models and parameter values. Our adopted cosmology is that of WMAP-7 \citep{Komatsu11}.    

Although performed at sufficiently high resolution to be able to resolve small
substructure such as satellite dwarf galaxies, the resolution is not
sufficient to resolve individual star formation regions or supernova
events. These physical processes are governed by the well-calibrated and
well-tested physics model used for the EAGLE project
\citep{Schaye15,Crain15}. Our implementation uses the parameter values
from the L100N1504 reference simulation presented by \citet{Schaye15}. The
star formation is modelled to follow the Kennicut-Schmidt star formation law. The rate of star formation is calculated depending on pressure rather than (more commonly used) volume density \citep{Schaye08} with a metallicity dependent density threshold
\citep{Schaye04}. Stellar thermal feedback is modelled as presented by \citet{DallaVecchia12}. Reionization (of hydrogen) is modelled to be instantaneous at z=11.5 following the prescriptions of \citet{Haardt12} and \citet{Wiersma09b}.  

The stellar evolution feedback processes responsible for chemical enrichment include winds from AGB stars, Type Ia supernovae, Type II supernovae, and winds from their progenitors \citep[for details see the descriptions in][]{Wiersma09a}. The supernova Ia rates from \citet{Schaye15} are shown to be in broad agreement with the observed evolution of the supernova Ia rate density. Yields are taken from \citet{Portinari98}, \citet{Marigo01}, and \citet{Thielemann03}. Each star particle is treated as a simple stellar population with a Chabrier IMF \citep{Chabrier03}. Finite stellar lifetimes are implemented that are metallicity-dependent \citep{Portinari98}. Stellar mass loss is distributed over 48 neighbouring particles using an SPH kernel transport scheme \citep{Pawlik08}, after setting the
mass of the gas particles equal to a constant initial value \citep{Schaye15}. Eleven different elements are followed throughout the simulation: H, He, C, N, O, Ne, Mg, Si, S, Ca, and Fe. The same elements are used to calculate radiative cooling. 

No special treatment of the first stellar
generation is implemented in our APOSTLE simulations, the first stars forming in the simulation are modelled with an identical initial mass function and supernovae feedback as later generations. However, even if the first stars should be very peculiar in their properties, star formation
processes will generally normalise in a few generations when metals become
available. We will therefore focus on the first generations of stars that
were formed, rather than the very first stars themselves. 

\section{The oldest and most metal poor stars}\label{sec:oldormetpoor}
In this work we focus on the most metal poor and oldest generations of stars and their distribution in the simulations. We define ``oldest'' and ``most metal poor'' as follows for the remainder of the paper:

\begin{itemize}
\item{\textbf{oldest:} formation time $< 0.8$~Gyr after the Big Bang (i.e., redshift $> 6.9$)}
\item{\textbf{most metal poor:} $[{\rm Fe}/{\rm H}] < -2.5$}
\end{itemize}

These choices are naturally somewhat arbitrary, but nevertheless motivated by
the results from observational surveys that tend to provide
spectroscopic results for stars in the low-metallicity regime from
$[{\rm Fe}/{\rm H}]=-2.5$ and below. The age restriction is motivated
by the approximate time scale for reionization. A time of 0.8~Gyr after the
Big Bang corresponds to a redshift of $z=6.9$, close to the
epoch of reionization\footnote{In our simulations reionization is forced to externally occur at $z$=11.5, this corresponds to 0.4 Gyr after the Big Bang}. We want to refrain from making the age cut too close to the Big Bang itself, because as motivated already in the introduction, we have no firm knowledge of the formation of the very first stars and no special modelling of these is provided in our simulations. We therefore believe our results will be more robust when the focus lies on the first several generations rather than the very first alone. An additional advantage of the cuts made is that the numbers of most metal poor stars and the numbers of oldest stars according to these criteria are roughly equal. We refer the reader to Section \ref{sec:disc} for a discussion on the robustness of our results regarding these choices.

\subsection{Metallicity and age distributions in the APOSTLE galaxies}\label{sec:metageapostle}

\begin{figure}
\includegraphics[width=\linewidth]{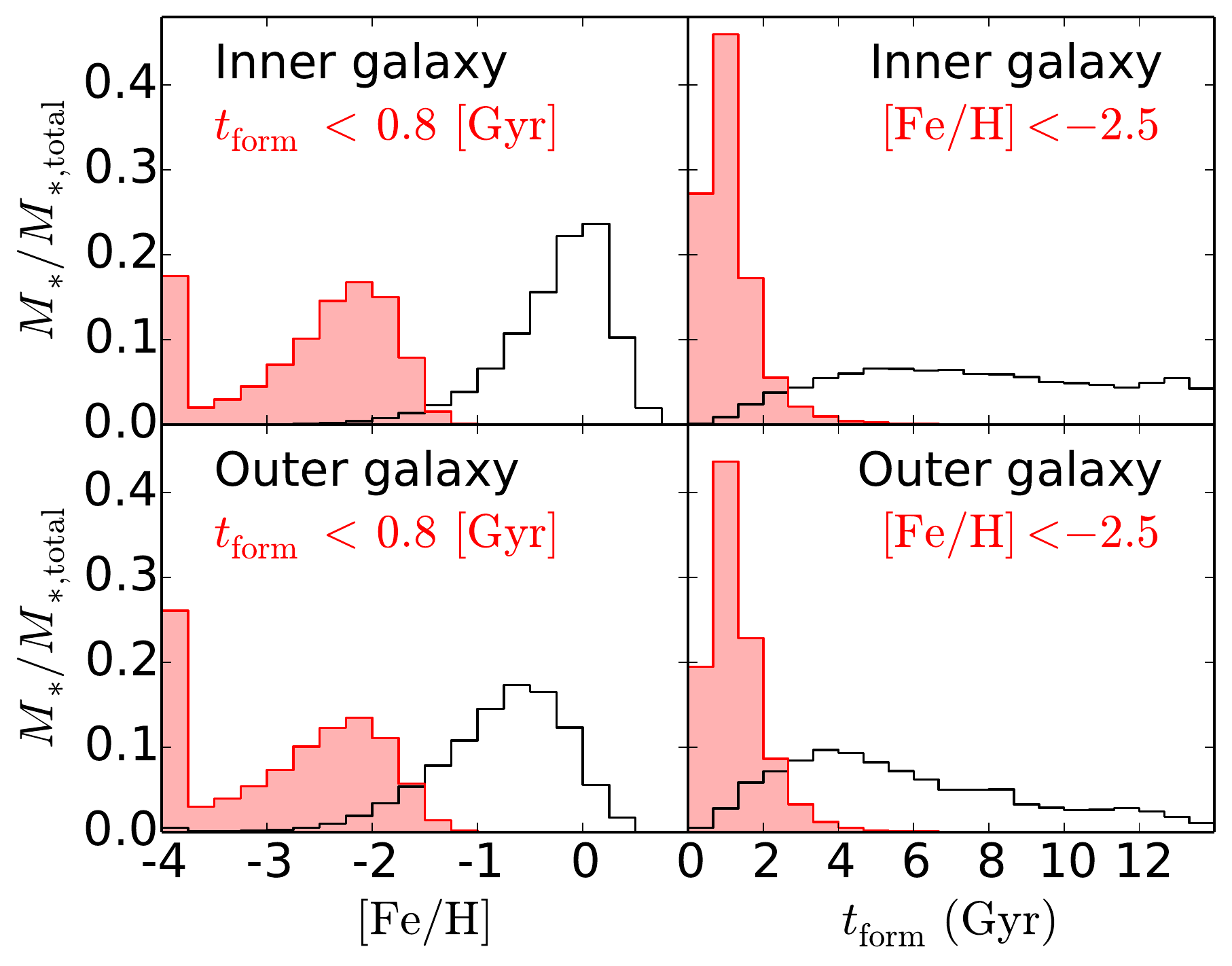}
\caption{Metallicity and age distribution functions for all APOSTLE main galaxies (resembling the MW and/or M31 in mass) in all intermediate resolution L2 runs.  We represent either the inner regions of the galaxies ($r<15$~kpc), or their outer halo populations ($15 < r < 100$~kpc), where for the latter we have left any in-situ populations out (see text for details). The open black histograms represent the full sample of all stellar particles within these radial cuts. The shaded red histograms represent a subset of the oldest stars (in the two left panels; formation time $< 0.8$~Gyr after the Big Bang) or the most metal poor stars (the two right panels; [Fe/H] $<$-2.5) for the same radial selections. Both the open and shaded histograms are normalised to the total mass in stars in their sample, the red shaded histograms contain in all cases less than 2\% of the stars in the open black histograms. Stars with metallicities less than -4 are shown in the lowest metallicity bin.}
\label{fig:agemethist}
\end{figure}

Before directing our attention fully to the oldest and most metal-poor stars, we first investigate the overall metallicity and age profiles of the APOSTLE main haloes in order to understand the main properties of the galaxies we study in this work. Open histograms in Fig.\ref{fig:agemethist} show the stacked age and metallicity distribution functions of the main galaxies (primary and secondary haloes of each pair) within the APOSTLE simulations for all intermediate resolution L2 runs -- these are available for all twelve volumes. In this analysis, we include any bound substructures that have survived in the main haloes. We represent metallicity by $[{\rm Fe}/{\rm H}]$, as is common in observational studies\footnote{In this notation $[{\rm Fe}/{\rm H}] = {\rm log}_{10}\bigl(\frac{N_{\rm Fe}}{N_{\rm H}}\bigr)_{\rm stars} - {\rm log_{10}}\bigl(\frac{N_{\rm Fe}}{N_{\rm H}}\bigr)_{\rm Sun}$.  If instead of $[{\rm Fe}/{\rm H}]$ we use $[{\rm M}/{\rm H}]$, regarding all metals rather than only iron, none of our main results are affected.}. Additionally, we have chosen to use the SPH-smoothed metallicity values. As described in Section 5.2 of \citet{Wiersma09a}, the smoothed metallicity formalism -- in which the metallicity of a star is set at its time of formation as the ratio of the SPH-smoothed gas metal mass density and the SPH-smoothed gas mass density -- has the advantage that it is better tailored to the nature of SPH simulations and furthermore that it somewhat alleviates the absence of a metal mixing formalism in the code. The smoothed abundances were also used in the simulations themselves for the calculation of cooling rates, stellar yields and stellar lifetimes. 

In Figure \ref{fig:agemethist}, a rough distinction is made between the inner and outer galaxy components, to be within or outside a radius of 15~kpc from the galaxy's centre
respectively (and within 100 kpc). SPH in combination with subgrid star formation prescriptions are known to be sensitive to the precise treatment of thermal instabilities and can sometimes form stars in unstable pockets of (initially hot) halo gas in larger galaxies \citep[e.g.,][]{Kaufmann06,Keres12,Parry12,Hobbs13,Cooper15}. Although the severity of this problem is greatly reduced in EAGLE by the use of the ``Anarchy'' hydrodynamics solver \citep{Schaller15}, we avoid any influence of these numerical ``nuisance'' events on our results, by showing in the panels, for stars at distances $> 15$ kpc, only stars that have not formed ``in situ'' in the halo of the main galaxy. In practice, this means that the outer galaxy only includes the ``accreted'' component of the stellar halo, as well as other bound substructures. 

While the build-up of the inner galaxy as well as the halo metallicity distribution function over cosmic time is interesting in its own right, we leave this as a topic for future work;  for the purposes of
this paper, it is mainly interesting to note that there is a difference between both populations in
terms of ages and metallicities. In the outer regions beyond 15 kpc, these galaxies have
a peak at earlier formation time ($\sim3$ Gyr) and lower metallicities ($[{\rm Fe}/{\rm H}] \approx -0.6$). Additionally, for the
outer components, the tail of the metallicity distribution
extends to lower metallicity. Qualitatively, this agrees with observational evidence in the Milky Way where we also see that the outer halo contains a much larger fraction of old, metal poor, and very metal poor stars when compared to the more central regions \citep[e.g.,][]{Helmi08,BlandHawthorn16}. However, the peak metallicity of the Milky Way halo population is at a lower value. \citet{AllendePrieto15} find that for their sample of tens of thousands of turn-off stars with $g$-band magnitude $>17$ and distances ranging from 5 to tens of kpc from the Sun, the peak of the metallicity distribution is at $[{\rm Fe}/{\rm H}] = -1.6$, dropping to $[{\rm Fe}/{\rm H}] \approx -2.2$ for stars beyond 10 kpc from the Sun \citep[these values are in good agreement with previous work][]{Carollo07,deJong10,Carollo10,Beers12}. Although we do see some individual galaxies in the simulation with a lower metallicity peak in the outer galaxy, our mass-averaged main Local Group galaxies as shown here bear more resemblance to observations in M31 \citep[e.g.,][]{Durrell04,Gilbert14} where a halo population peaking at $[{\rm Fe}/{\rm H}] \sim -0.5$ is seen (even at a radius of 30~kpc or beyond), with a long tail towards low metallicities.  

As illustrated in the shaded histograms in Fig.~\ref{fig:agemethist}, the oldest stars in a galaxy tend to be more metal poor than the average and vice versa -- the most metal poor stars in a galaxy are older on average. However, within these shaded histograms there is clearly still a large range. Not all stars that are older than 0.8~Gyr are very metal poor (indeed, the population reaches its peak a bit below $[{\rm Fe}/{\rm H}]=-2$) and using our most metal poor cut we select stars with a range of ages up to several Gyr after the Big Bang. We find that 50\% of the most metal poor stars form within 1.1 Gyr and 90\% within 2.4 Gyr (corresponding to $z=5.3$ and $z=2.8$ respectively).

\section{Are the most metal poor stars also old?}\label{sec:oldandmetpoor}

\begin{figure}
\begin{center}
\includegraphics[width=\linewidth]{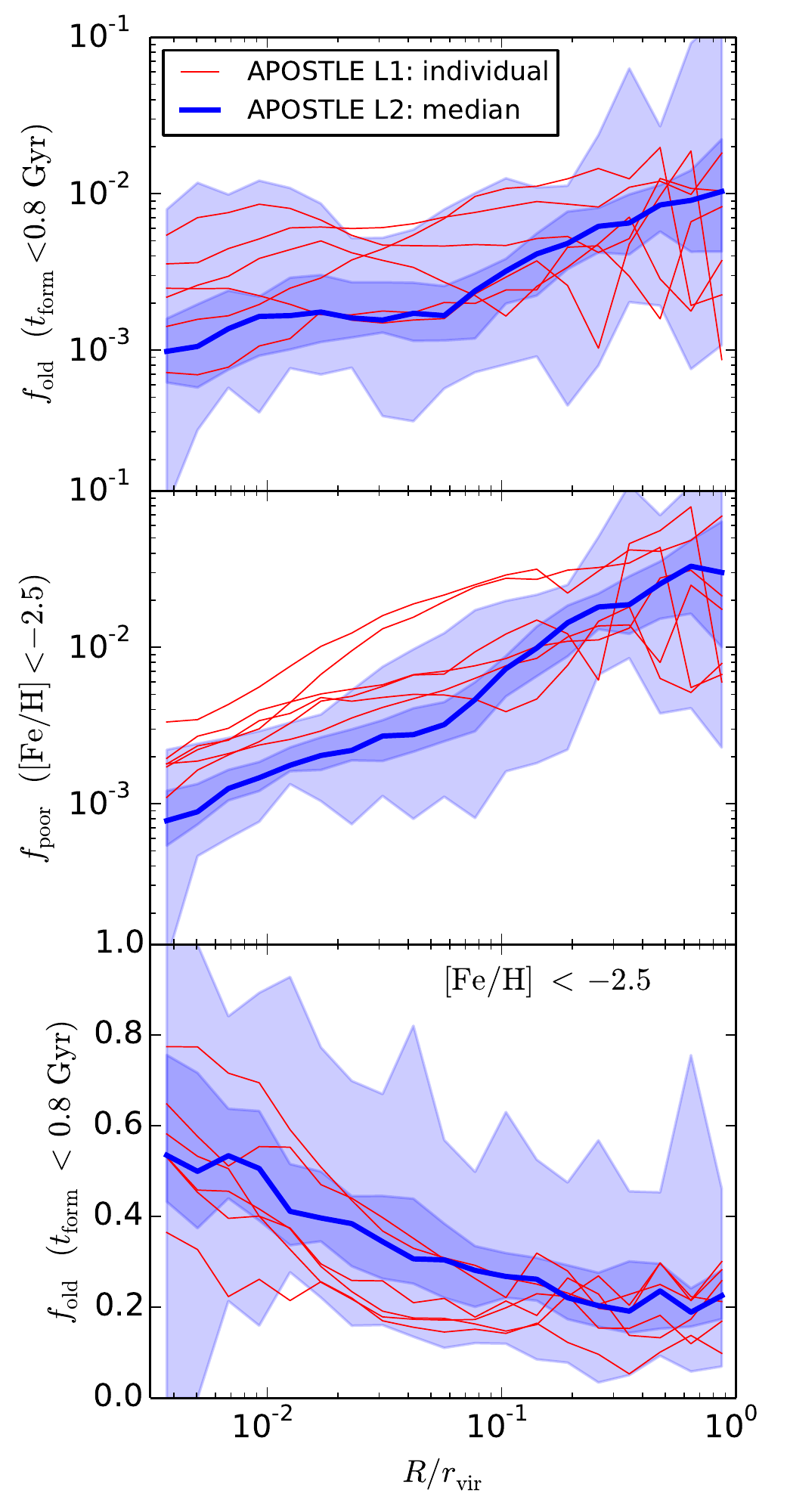}
\caption{Top and middle panels: the fraction of oldest and the fraction of most
  metal poor stars respectively as a function of radius in each of the main
  galaxies out to their virial radius. For the L2 simulations a median curve of all main galaxies
  is shown, whereas the six main galaxies simulated at higher resolution, L1, are shown as individual (red) lines. Blue shaded regions show the interquartile and full range of the L2 simulations. Bottom panel: the fraction of most metal poor stars that are old (see text for definitions).}
\label{fig:fR}
\end{center}
\end{figure}

In the previous section we have investigated the oldest and the most metal poor stars in the simulations; here we combine the two criteria and ask: if we observe a metal poor star in any particular environment, will it also be old? In other words: where will the first order approximation best hold that the most metal poor stars are messengers from the early Universe? This question is particularly relevant for observational surveys aiming to find metal poor stars and measure their metallicity, but that are unable to measure their precise ages. 

The top and middle panels of Fig.~\ref{fig:fR} show the fraction of oldest and the fraction of most metal poor stars as a function of radius in each of the main galaxies out to their virial radius. The fraction of both old and most metal poor stars increases in the outskirts of these galaxies, suggesting that a perfect place to look for either of these populations would be the Galactic halo. The bottom panel of Fig.~\ref{fig:fR} shows the radial profile within the main
galaxies, similarly to the top panels, but now for the most metal poor stars that also also old. We see that the trend is the reverse from that shown in the top two panels.
When a star is selected that is metal poor (as one
would in an observational study), the probability that that star is old is lower in the galaxy outskirts than in the centre. This general trend of
finding the oldest stars (among the metal poor stars) mainly in the centre of
the galaxy has been discussed before \citep{White00, Brook07, Gao10, Tumlinson10, Salvadori10}
and is generally explained by the first stars forming in the density
peaks that collapse first in the Universe, these then generally end up in the centres of the
main haloes. However, quantitatively the results of the various studies vary significantly, reflecting uncertainties in the modelling of early Universe physics. We refer the reader to the discussion in Section \ref{sec:disc} for a
more detailed and quantitative comparison of earlier work with our results.

The fractions shown here will naturally depend on the implementation of
star formation, feedback processes, and reionization in the
simulation. A weak trend of the fractions of old and most metal poor stars with resolution can also be seen in Fig.~\ref{fig:fR}; at small radii, most of the individual L1 runs deviate somewhat from the median result for all L2 galaxies (most notably in the middle panel).  In Fig.~\ref{fig:reseffects} in Appendix~A we demonstrate further that the populations we
focus on are only weakly affected by resolution effects, at least when comparing
the L1 and L2 runs. We will therefore focus on these resolutions for the remainder of the paper.
 
We additionally note that our results are virtually insensitive to the inclusion or exclusion of the possibly spurious in-situ outer halo star population discussed in Section \ref{sec:metageapostle}. While the fractions of old \textit{or} metal poor populations shown in the top panels of Fig.~\ref{fig:fR} show an increase when only accreted stars are considered, there is very little difference in the ratio of most metal poor stars that are old, as defined in the bottom panel. The reason is that in-situ stars form later and out of metal-enriched gas. Hence they increase the total population of stars, but do not affect the populations of interest in this paper.

\begin{figure*}
\raggedright
\includegraphics[width=0.95\linewidth]{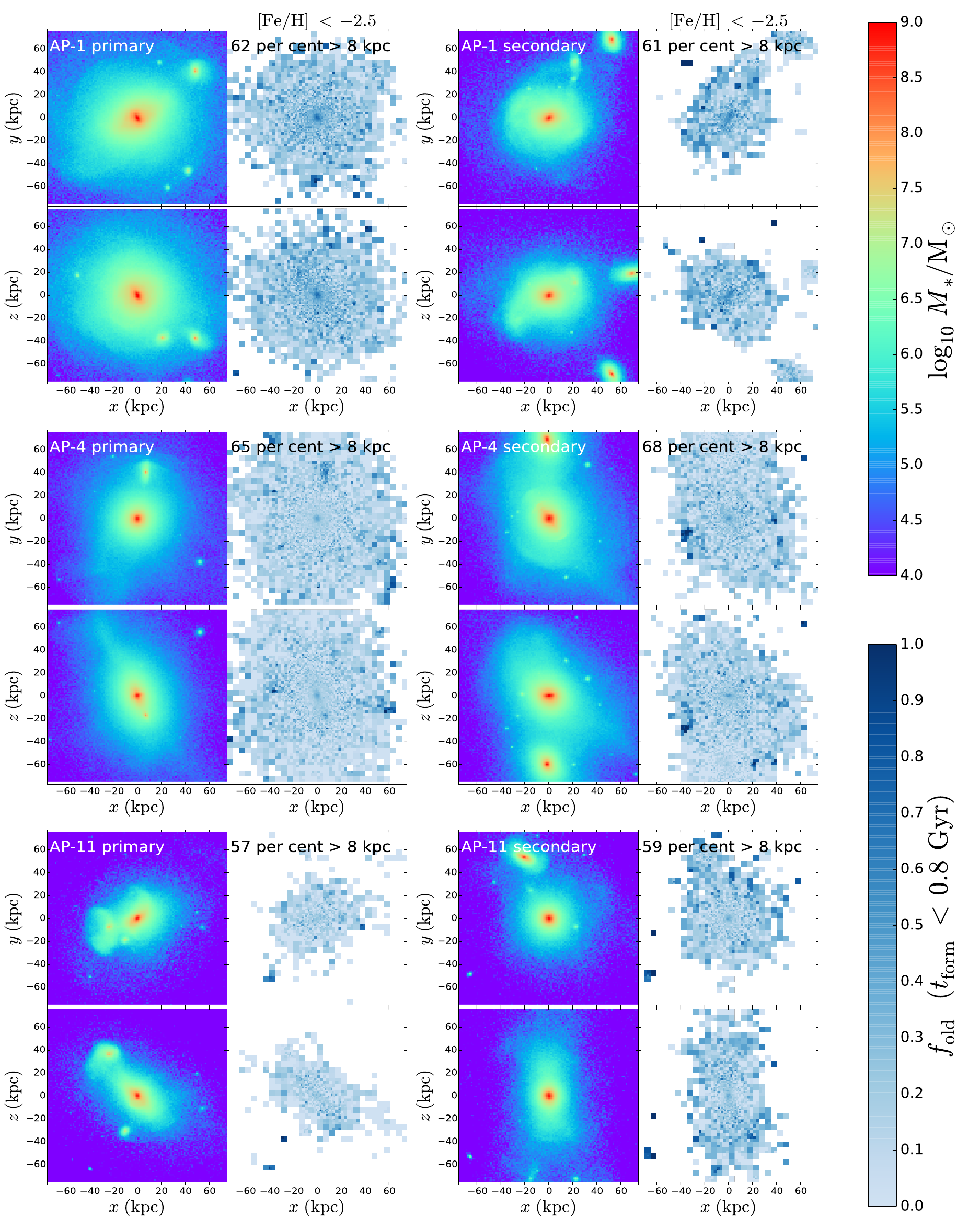}
\caption{Left panels: logarithmic stellar masses of the main galaxies
  in the high-resolution L1 simulations in $x$-$y$ and $x$-$z$ views in the coordinate frame of the simulation (the coordinates are not redefined based on the galaxy morphology). From top left to
  bottom right: AP-1 primary halo, AP-1 secondary halo, AP-4 primary halo, AP-4
  secondary halo, AP-11 primary halo, AP-11 secondary halo. The corresponding colour bar is given at the top right. Right panels: for each left panel, we show the most metal poor population in each spatial pixel colour coded by the fraction of old stars in this population (the corresponding colour bar is on the bottom right). The pixel size varies with density, each pixel has a population of at least 5 star particles, and lower density regions are not shown. The percentage shown at the top left of each of these panels indicates what mass fraction of these most metal poor \textit{and} old component are at distances larger than 8 kpc from the Galactic centre.}
\label{fig:XY}
\end{figure*}

Following up on the general trend of more centrally concentrated old \textit{and} metal poor populations, we explore the spatial distribution of the most metal poor populations in the galaxies in more detail in Fig.~\ref{fig:XY}. Shown here are two orthogonal projections of all stars of each of the six main galaxies in the three highest resolution L1 APOSTLE simulations. The left panels of each grid show a logarithmic colour coded stellar mass map of the galaxy out to 75~kpc from the galaxy centre. The right panels of each subplot show only those stars that are most metal poor according to our criterion in each spatial bin (the bin size is allowed to vary according to the density of stellar populations in that area), and the colour coding is chosen so that it follows the fraction of the most metal poor stars that are old within this population. Several striking features stand out from these figures:

\begin{itemize}
\item{We still see the general trend that the metal poor population is older at the center than in the outskirts, as was shown already in Fig.~\ref{fig:fR} and can be seen here by a darker shade in the centres of the galaxies in the right panels (most notably visible in AP-1 primary, AP-4 primary and secondary and AP-11 secondary galaxies). }
\item{However, in the outskirts of the galaxies,  typically at scales of several tens of kiloparsecs \textit{each} of our six modelled galaxies shows particularly interesting substructures that are on average very old -- frequently even more so than the metal poor stars in the central few kiloparsecs. The distribution of this population changes from galaxy to galaxy.}
\item{As can be seen from the comparison with visual overdensities, this population is often connected to their individual merging histories. The oldest substructures frequently correspond to bound or disrupting satellites. For instance, all darkest pixels in right panels of the AP-4 secondary galaxy, or AP-11 primary galaxy correspond to compact overdensities in the stellar density maps. In the AP-1 secondary galaxy we see a large satellite galaxy disrupting at $(x,y) \approx (20,50)$. Careful inspection of the stellar density map reveals a bridge of stars connecting it to the main host (from $(x,y) \approx (20,50)$ to $(x,y) \approx (-20,30)$). We observe the same feature in darker shades in the right panel.}
\item{These metal poor \textit{and} oldest populations in the halo generally follow substructures, but not \textit{all} substructures. More importantly, they sometimes trace merging events that are not apparent in the overall stellar density because the metallicity and age contrast is much greater than the surface density contrast. An example of this can be seen in the secondary halo of AP-1 (top right panel), where at $(x,y) \approx (0,-55)$ a very old and metal poor substructure stands out that is not visible in the full stellar density map.}
\end{itemize}

\begin{figure*}
\begin{center}
\includegraphics[width=\linewidth]{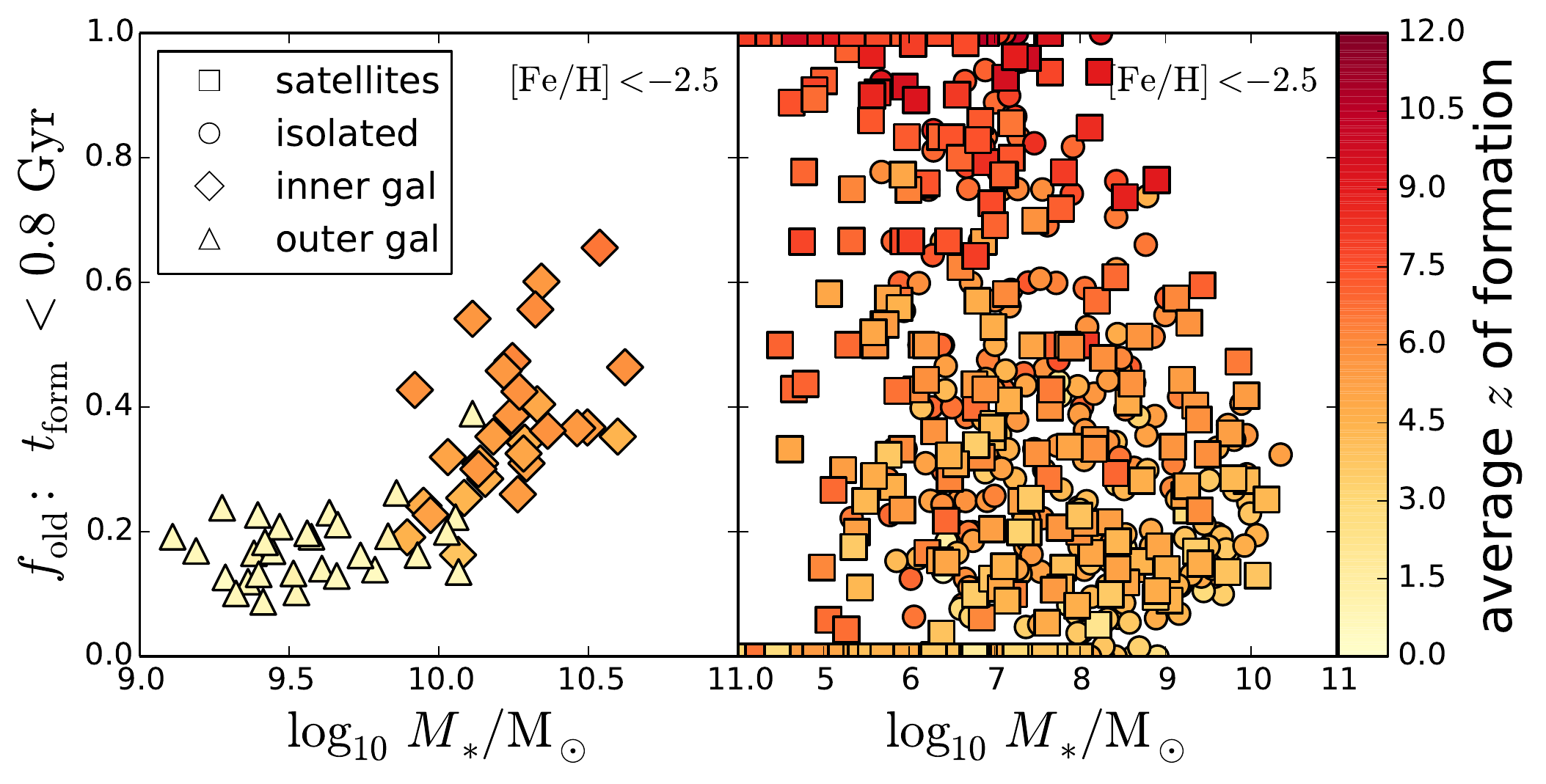}
\caption{The left panel shows the mass fraction of old stars among the most metal poor populations shown
  for the inner and outer galaxy components (defined as $r < 15$~kpc and $15 < r < 100$~kpc, shown as diamonds and triangles,
  respectively) of each main galaxy as a function of the total stellar mass at those radii. The right panel shows
  the same metric, but now for every dwarf galaxy included in the
  simulation within a 2~Mpc sphere. Satellite systems, within 300~kpc of either one of the main galaxies, are shown as squares, isolated dwarf galaxies as circles. The symbols are colour coded according to the average redshift of formation of their most metal poor population. The colour gradient that is visible across the y-axis of the figure confirms that the metric chosen here (the fraction of most metal poor stars that are also old, where old means formed before 0.8 Gyr or a redshift of 6.9) is indeed correlated with the average formation time and that no spurious behaviour is introduced by imposing a sharp threshold for what we call ``old''. \label{fig:foldpoorsatmains}}
\end{center}
\end{figure*}

In summary, these maps show that even though the average age of metal poor stars decreases in the outskirts, there are regions connected with bound or disrupted substructures where the chance of tracing early Universe physics in the population of metal poor stars is exceptionally high. To quantify this result, we have labelled each of the panels with a percentage indicating the mass fraction of these most metal poor \textit{and} oldest stars that are found outside the solar circle (at distances $>$ 8 kpc, the total sample is defined to be within 300 kpc of the galaxy's centre). For each of the galaxies shown this mass fraction is $\sim$60 per cent, indicating that more of these interesting stars are found outside the solar radius than within\footnote{Naturally, not each of these galaxies represents a Milky Way and the shapes and sizes of their stellar components vary, which means that we are not probing the same mix of stellar components at this fixed radius \citep{Campbell16}.}. We verified that this result is robust to the simulation resolution; in the medium resolution L2 galaxies we also find very similar fractions for the most metal poor and oldest stars outside the solar radius, with a median for all 24 galaxies of 62.3 per cent. The large majority of these most metal poor and oldest stars outside the solar radius are part of the diffuse stellar halo (i.e., they do not belong to any bound substructure at $z=0$), but nevertheless a significant fraction (on average 19 per cent) of them belong to bound substructures. Approximately half of the population of most metal poor and oldest stars is found outside a radius of 15 kpc, indicating that surveys need to reach out quite far to study the full population. To observe the sample to 90\% completeness, one has to extend the survey out to $\approx$ 95 kpc. In the next section we look into the various surviving substructures to see if we can disentangle the systems that stand out in this respect from their peers.

\section{Comparing the halo population to the surviving satellites and isolated dwarf galaxies}\label{sec:dwarfs}

In Fig.~\ref{fig:foldpoorsatmains} we investigate the fraction of oldest stars among the most metal poor stars for galaxies as a whole. In this comparison, we include all galaxies with a stellar mass over $10^4$M$_{\odot}$, comparable to the Milky Way satellites Bo\"{o}tes I, Leo IV, or Leo V which are on the brighter end of the so-called ultra-faint galaxies. While we can include also more isolated dwarf galaxies down to this mass limit, these will be typically too faint to be observed; Leo T at a distance of over 400 kpc that has $\approx$ 1.4$\times10^{5}$ M$_{\odot}$ is at the low luminosity end of the more isolated dwarf galaxies \citep{McConnachie12}. Like in observational data, we find that our simulated dwarf galaxies obey a mass-metallicity relation where the fainter galaxies are more metal poor. For a typical dwarf galaxy of $10^{6}$ M$_{\odot}$, comparable to the Sculptor dwarf galaxy, we find that its average metallicity is $\approx$-1.6 in agreement with the measured metallicity distribution function of the Sculptor dwarf spheroidal galaxy \citep[e.g.,][]{Romano13}.  For the main galaxies we again distinguish the inner and outer halo regions, in a similar fashion as in Fig.~\ref{fig:agemethist}. As can be seen in the left panel of Fig.~\ref{fig:foldpoorsatmains}, for the inner Galaxy component of a Milky Way-mass galaxy the fraction of the most metal poor stars that are old is typically 0.3 -- 0.4, whereas for the outer Galaxy this fraction is typically 0.1 -- 0.25. However, dwarf galaxies (shown in the right panel of Fig.~\ref{fig:foldpoorsatmains}) span the full range from 0 to 1. In the right panel we include both the satellite galaxies that today reside close to the main hosts ($<300$~kpc, shown as squares), as well as more isolated systems (shown as circles). Both populations span the whole y-axis and we find no systematic differences between them. It should be noted here that we do not specifically focus on the very tiny galaxies or ``fossil galaxies'' mentioned in the introduction. Many of
those will fall below our resolution limit even at our highest resolution. Instead, we are
concerned with the low-metallicity population of any dwarf galaxy above our
resolution threshold (no matter
what its peak metallicity or star formation history is) and we investigate
whether its metal poor population is also old. 

If we could isolate the oldest stars in these systems, then this would offer great potential for early universe studies across various environments. However, as the right panel of Fig.~\ref{fig:foldpoorsatmains} illustrates, any given star of $[{\rm Fe}/{\rm H}]<-2.5$ in a dwarf galaxy system might be a messenger from the distant past, but could also be a much more recently formed object. How can we distinguish between those satellite systems that host stars formed before or during the epoch of reionization and those that don't? 
\subsection{How to find the systems where more metal poor stars are old}\label{sec:howto}

\begin{figure}
\begin{center}
\includegraphics[width=\linewidth]{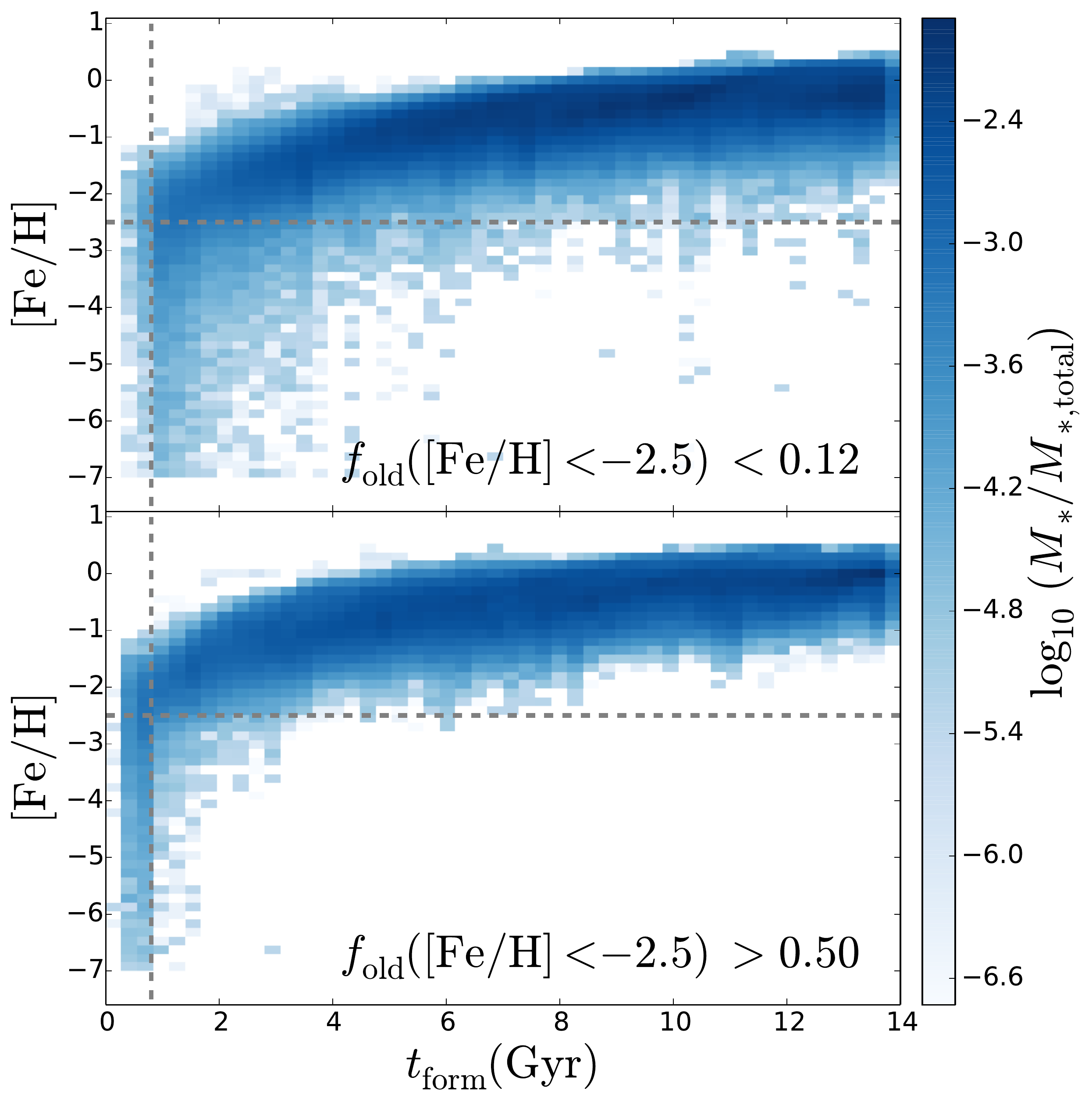}
\caption{Age-metallicity relations for dwarf galaxies that have a high fraction of oldest stars among their most metal poor populations (bottom panel) and those that have a low fraction (top panel). The fractions are chosen so that the sum of their stellar masses in both panels is roughly equal. Grey lines indicate the definitions of oldest and most metal poor as defined in Section \ref{sec:oldormetpoor} and used here and throughout this paper.}
\label{fig:agemetsats}
\end{center}
\end{figure}

\begin{figure}
\begin{center}
\includegraphics[width=\linewidth]{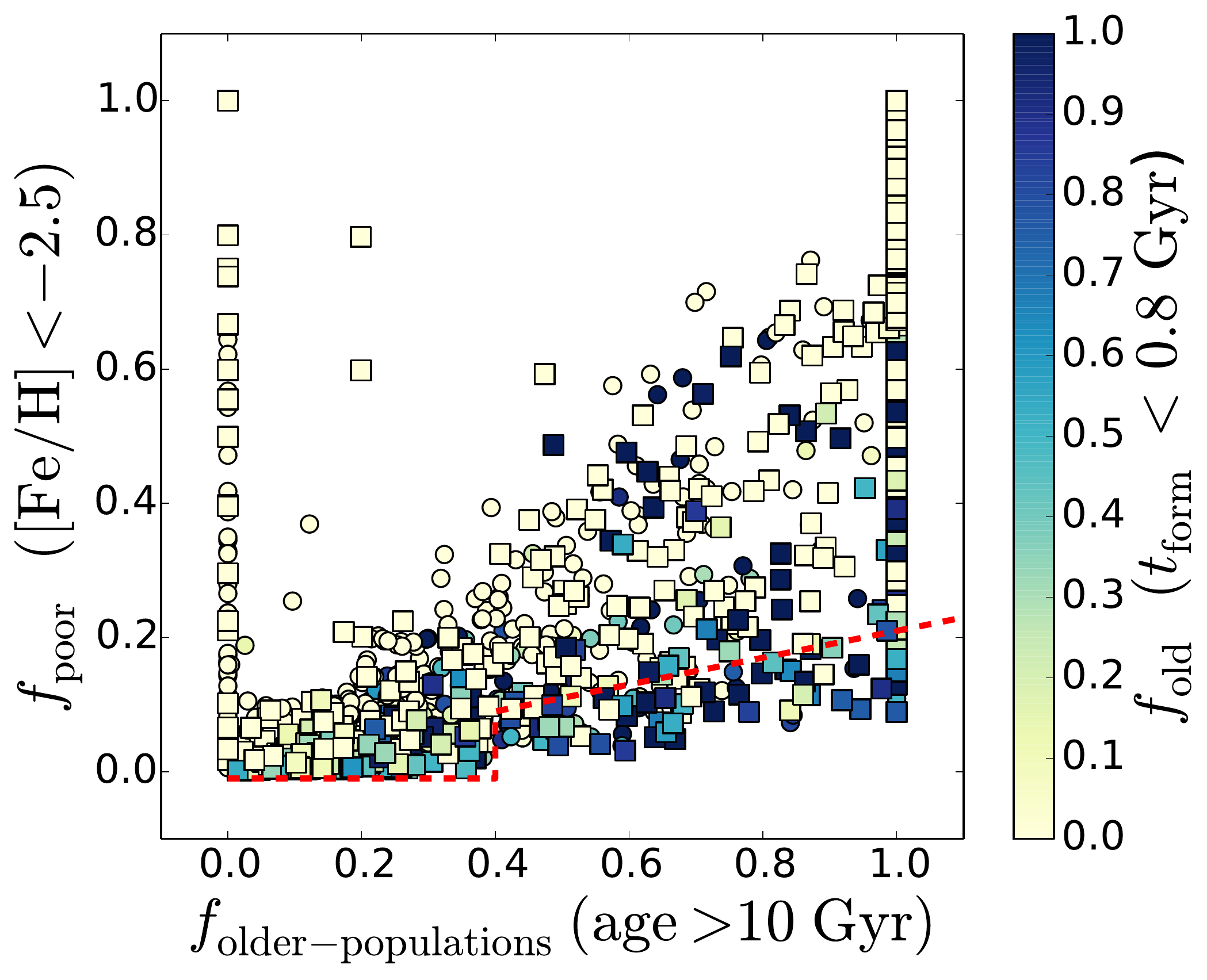}
\caption{The fraction of age $> 10$~Gyr populations in dwarf galaxies and the fraction of stars with $[{\rm Fe}/{\rm H}] < -2.5$, colour coded by the fraction of these most metal poor stars that are old (formed before 0.8 Gyr after the Big Bang). Squares are satellite galaxies within 300~kpc of their host, circles indicate more isolated systems. The red line indicates a suggested cut in this parameter space (sloping up from $< 10$ to $< 20$ per cent most metal poor stars of their total stellar population, as a function of the fraction of stars older than 10 Gyr) to select the systems with a high chance of having very old stars amongst their most metal poor population (see text for details).}
\label{fig:foldpoorobs}
\end{center}
\end{figure}

In Fig.~\ref{fig:agemetsats} we analyse the age-$[{\rm Fe}/{\rm H}]$ distributions for those dwarf galaxies that have a high fraction of old stars among their most metal poor populations (in the bottom panel) and those that have a low fraction (in the top panel). It is clear from this figure that the dwarf galaxies with high fractions of old stars among their most metal poor stars start to form stars earlier \textit{and} are more effective at enriching themselves; they do not form metal poor stars anymore at later stages.

It is in the interest of future surveys hunting for old stars to separate the galaxies in the bottom panel of Fig.~\ref{fig:agemetsats} from those in the top panel. However, it is very challenging observationally to provide the necessary accuracy on the age determination. In general, it is difficult to find any imprints of the very first stages of star formation still in the dwarfs today. For most dwarf galaxies, the overwhelmingly larger more metal-rich and younger populations contain little information on the first stages of star formation -- indeed, we found no clues to the value of f$_{\rm old}$ in any present day observable like star formation history, metallicity distribution or even morphology or compactness of the resulting dwarf galaxy. 

One interesting parameter that can be measured, however, even in distant systems, is the fraction of the stellar population older than $\sim$10~Gyr (because this population has specific tracers, i.e., RR-Lyrae stars). In Fig.~\ref{fig:foldpoorobs} we show how this metric, if combined with a measurement of the (mass-)fraction of the most metal poor stars, can provide a much more successful identification of dwarf galaxy systems in which to search for very old stars. 

Without any preselection we find that in $\sim$16 per cent of the dwarf galaxies observed (both field and satellite galaxies) over 50 per cent of their most metal poor stars are old. Since these are the systems most interesting for old star search campaigns, we define our blind search ``success rate'' as 16 per cent. However, if instead of selecting all dwarf galaxies, we only select systems that have a high fraction of their population that is older than 10 Gyr ($> 40$ per cent) and a \textit{lower} fraction of metal poor stars (as illustrated by the red dashed line in Fig.~\ref{fig:foldpoorobs}) we increase this success rate to 57 per cent.

Our finding that one is better off selecting systems with very few metal poor stars in order to find the systems with the oldest stars, is somewhat counter-intuitive, but can be explained as follows: these systems had a very rapid evolution early on -- they enriched themselves quickly and did not form many metal poor stars at later epochs. Therefore, their present day populations are dominated by more metal-rich stars.

\begin{figure*}
\raggedright
\includegraphics[width=\linewidth]{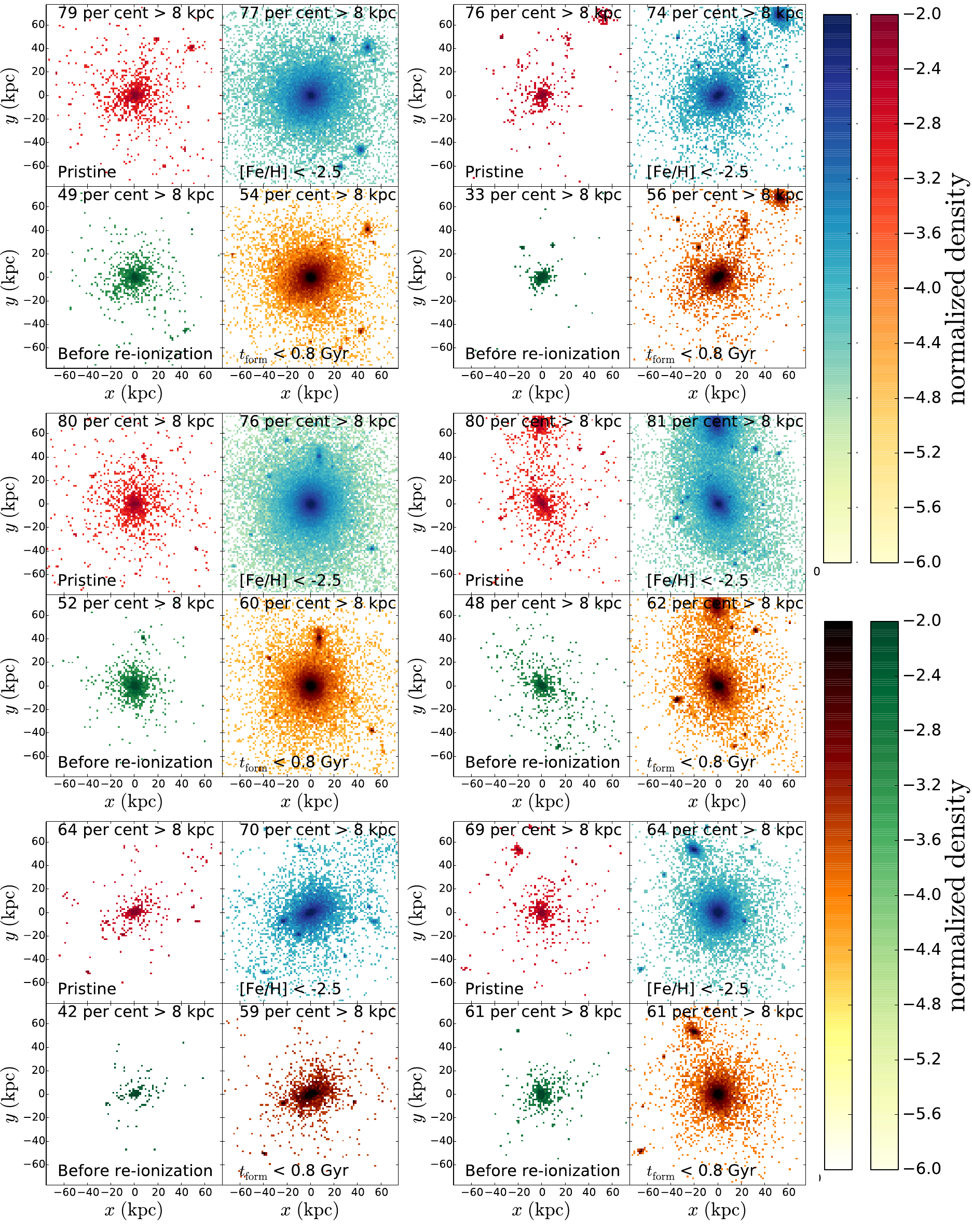}
\caption{Four different views on the six high-resolution main galaxies. Shown are the pristine population, completely devoid of metals (top left), $[{\rm Fe}/{\rm H}] < -2.5$ metal poor population (top right), total stellar population formed before reionization (bottom left) and the old population ($t_{\rm form} < 0.8$~Gyr after the Big Bang, bottom right). The percentage shown at the top of each of these panels indicates the mass fraction of these stars with distances larger than 8 kpc from the Galactic centre.}
\label{fig:XY4}
\end{figure*}

\section{Discussion}\label{sec:disc}

\subsection{Where are the first stars?}

In Fig.~\ref{fig:XY4} we again show a view of our six high-resolution main galaxies. The right panels of each 2$\times$2 grid show the oldest and most metal poor populations separately. On the left we show additionally a view of the stars formed before reionization in our simulation and of the perfectly pristine stars (those that are unpolluted by any metals). As discussed in the introduction, we do not necessarily believe that these stars should still survive in the simulation (they might have a very different IMF), nor that we are appropriately set up to trace these populations. Nevertheless, we show them here to illustrate and discuss how the sites in our simulations where the first and pristine star formation has occurred are distributed in the present day, as they might also trace the sites where we can find second generation stars that show their chemical imprint. In all galaxies, the pristine population has a total mass of only 2-5 per cent of that of the most metal poor stars, a negligible fraction. However, their radial distribution shows great similarity to that of the most metal poor stars, as can also be seen from the percentage of each of the populations found at radii beyond 8 kpc.

Similar to Fig.~\ref{fig:XY} we have labelled each panel in Fig.~\ref{fig:XY4} with a percentage indicating the mass fraction of these stars that are found outside the solar circle (at distances $>$ 8 kpc). The results clearly show that in our simulation the outskirts of the galaxies, beyond the solar radius, do not only contain metal poor and old stars, they also contain a significant number of pristine stars and stars formed before reionization (corresponding here to a redshift of 11.5). A similar conclusion can be reached for the smaller still-bound satellites, some of which contain pristine stars that were formed before reionization. It is interesting to note that there is significant variation from galaxy to galaxy. Additionally, we still see a trend showing an even more centrally concentrated population of pre-reionization stars compared to pristine stars. Of the latter we find that 75 per cent is found at larger radii than 8 kpc, whereas the former only has 50 per cent of its population at those radii. If we focus solely on even older stars, formed at $z >15$ (not shown), we find their population to be even more centrally concentrated. However, also here we find that typically at least a few percent of them reside at radii $>$ 8 kpc today, only in one galaxy all stars are contained within 8 kpc.

Based on these results we conclude that surveys primarily interested in finding the \textit{oldest} stars are still best served by focussing on the central Galaxy, although this has of course to be balanced with the difficulties of observing those regions. Observationally, the satellites of the Milky Way are much more accessible: many are relatively close, often visible at
high Galactic latitude, and reside in the low-density Galactic halo environment. We find that surveys interested in studying first star formation in various environments as well as very old stars are also well-off targeting the regions outside the solar radius and the Galactic halo with its streams and bound substructures. They are very likely to still uncover first stars, or, if these stars have not survived, their chemical imprints on the next generations.

\subsection{Comparison with other work}
As mentioned in Section \ref{sec:oldandmetpoor}, our finding that the stars that are both metal poor \textit{and} old are centrally concentrated agrees qualitatively very well with the
conclusions reached by earlier work
\citep{White00, Scannapieco06, Brook07,Tumlinson10, Salvadori10, Gao10}. Quantitatively, however, there are many
differences. 

Because of the great advances in hydrodynamical simulation techniques and computational power, we can work here at much higher resolution than was possible before \citep[for instance when compared to][]{Brook07} and without having to resort to semi-analytical techniques used to tag stars in N-body dark matter simulations \citep[as in][]{White00,Scannapieco06,Tumlinson10,Salvadori10,Gao10,Cooper10}. 

\citet{Scannapieco06} and \citet{Brook07}  explore the distributions of the first and second generations of stars with a semi-analytical technique grafted on an N-body simulation with a resolution similar to our L2 simulations, and subsequently with a resolution several order of magnitudes lower with a hydrodynamical technique. \citet{Scannapieco06} draw several conclusions that are robust to the particular metal dispersion prescription and other approximations in the semi-analytical code: in particular they find a population of metal free stars inside the Galactic halo environment that formed as late as redshift $z=5$ (or in some cases even $z=3$) and that extends to very large radii. \citet{Brook07} confirm these findings and show how also in their models the oldest stellar populations are more centrally clustered, but metal free stars can still be found at large radii. 

 Comparing these and our results with the \citeauthor{Tumlinson10} model, one striking difference is that  in \citet{Tumlinson10} \textit{all} stars with $[{\rm Fe}/{\rm H}]<-2.5$ are formed before $z=6$ (i.e., all their most metal poor stars are old according to the definitions used in this paper) and in
their fiducial model with the epoch of reionization at $z=10.5$, all stars
with $[{\rm Fe}/{\rm H}]<-3$ form before $z=10$. Their radial gradient result of more
centrally concentrated older stars thus merely differentiates  between $z=10$ and $z=15$
at most, corresponding to $\sim200$~Myr. Just like in our simulations, \citet{Tumlinson10} does not adopt any
special IMF or formation mechanism for the first generations of stars. However, in our chemodynamical model, clearly many stars with low metallicities are formed later than
$z=6$. Specifically, \citet{Tumlinson10} show that in their model, the fraction of extremely metal poor stars (selected as $[{\rm Fe}/{\rm H}] < -3.0$) that are born before $z = 15$ changes from $\approx$13 per cent in the inner regions to just a few per cent in the outer regions. When we make identical cuts in our simulations we find that a very low percentage of stars is formed at $z > 15$, even at such low metallicities. In the L2 galaxies, none are formed at all, indicating (as expected) a dependence of resolution on the very first star formation in the simulations. In the L1 galaxies, we always find the ratio of $[{\rm Fe}/{\rm H}] < -3.0$ stars that are formed at $z > 15$ to be below 3 per cent. No radial gradient is apparent for this fraction. 

\citet{Salvadori10} find that while the relative contribution of very metal poor stars increases with radius from the Galactic centre, the oldest stars populate the innermost region, in agreement with the conclusions reached in our work. In the fiducial model of \citet{Salvadori10} an instantaneous and homogeneous mixing prescription is used to mix metals within each halo's gas reservoir, and similarly also within the diffuse gas in the Milky Way for metals that are blown out by supernova explosions. This results in a pre-enrichment of the diffuse gas in the Milky Way up to $[{\rm Fe}/{\rm H}]=-3$ by a redshift of 7. They report that due to this prescription no stars with metallicities lower than $[{\rm Fe}/{\rm H}]=-3$ are found in  80 per cent of Milky Way satellite galaxies, a prediction that disagrees with our findings and that we now know to be contradicted by observational evidence \citep[e.g.,][]{Kirby08, Starkenburg10,Tafelmeyer10,Frebel10,Kirby10,Venn12,Starkenburg13,Frebel14,Jablonka15}. Their experiment with an inhomogeneous mixing prescription does not affect the simulation results significantly, but they stress that the true mixing at low metallicities or large radii remains very uncertain.

In \citet{Gao10}, molecular hydrogen cooling in minihaloes is modelled as a recipe on top of the dark matter merger tree backbone of the Aquarius simulations \citep{Springel08}. The explicit cooling of molecular hydrogen requires very high resolution and strong assumptions on the photodissociating background radiation and is hence ignored in all other work mentioned here, including our own. They find the minihaloes to be strongly clustered in space, therefore limiting the number of final building blocks that contain them as most of them merge in time. At the present day, the environments where the first stars are found are again centrally concentrated, but they also report that a significant fraction (over 20 per cent, a very similar percentage as found in this work as discussed in Section \ref{sec:oldandmetpoor}) remains in the satellites that are presently surrounding the Milky Way. In \citet{Gao10}, none of the baryonic processes are modelled after the first generation of stars, so the present day properties of these satellites can only be compared on the basis of their dark matter properties, which, however, do seem to be comparable to the surviving satellites that we observe around the Milky Way today.   

\subsubsection{The role of metal mixing}

\begin{figure}
\includegraphics[width=\linewidth]{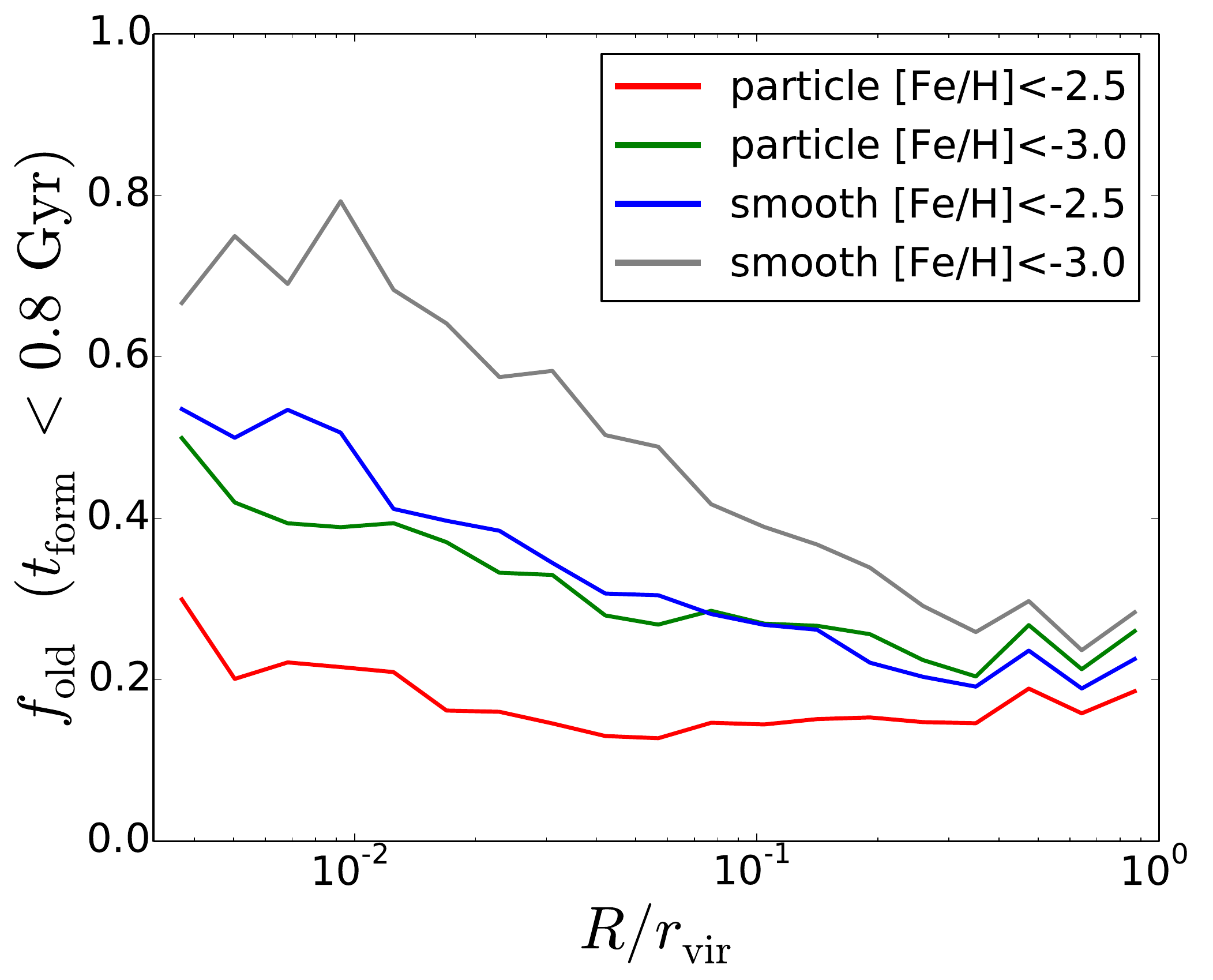}
\caption{The fraction of most metal poor stars that are old as a function of radius. Shown here is the median value for all L2 galaxies (corresponding to the blue thick line in the bottom panel of Fig.~\ref{fig:fR}; the blue line in this figure is identical to that) for two different metallicity calculations: an SPH-smoothed metallicity (our fiducial choice) and a particle metallicity (no metal mixing at all). We additionally show two different cutoff values for the most metal poor stars: $[{\rm Fe}/{\rm H}] < -2.5$ (as used throughout the paper) and $[{\rm Fe}/{\rm H}] < -3.0$.}
\label{fig:recipes}
\end{figure}

The many differences between the models once more illustrate how the results at early epochs and low metallicities are sensitive to the assumptions that are made and the modelling prescriptions that are used. Here we investigate in more detail in particular the sensitivity of our results to the assumptions on metal mixing. As described in Section \ref{sec:metageapostle}, our fiducial model includes an SPH kernel-smoothed mixing prescription. However, in the APOSTLE simulations we also store the individual star particle metallicities, inherited from unmixed gas. In general, we find that all qualitative results are robust to either calculation of the metallicity and that most quantitative results show only very minor variations (for instance, the percentages of the different populations that are located outside an 8 kpc radius as labelled in Figures \ref{fig:XY} and \ref{fig:XY4} are robust to the metallicity prescription used). 

However, there are significant changes in the radial dependence of the fraction of the most metal poor stars that are old. This is illustrated in Fig.~\ref{fig:recipes} where the median fraction for all L2 galaxies is shown as a function of radius for both types of metallicity estimates (the blue line corresponds to the blue thick line in the bottom panel of Fig.~\ref{fig:fR}). Fig.~\ref{fig:recipes} additionally shows that there is a degeneracy between the amount of mixing and the cutoff  $[{\rm Fe}/{\rm H}]$ used to determine whether a star is metal poor. For the particle metallicity scheme we still see a clear gradient when we adopt a lower metallicity threshold. Whereas SPH-based results will generally underpredict mixing and use particle metallicities, semi-analytic models often assume perfect mixing of all gas within a system. Our implementation of smoothed metallicities lies in between these two extremes, but while it decreases the severity of the sampling aspect of the mixing problem within SPH, it does not solve it either \citep{Wiersma09a}. Clearly, quantitative results from any prescription quoting the fraction of metal poor stars that are old in a certain Galactic region should be treated with caution. They are very dependent on the assumptions for metal mixing, a process that is uncertain.

\section*{Conclusions}

We have investigated the properties of the most metal poor and oldest stellar populations in the APOSTLE simulations of Local Group analogues, focussing in particular on their present day distribution. As illustrated in Fig.~\ref{fig:fR}, we find that the highest fraction of either of these populations can be found at larger radii, making the outer halo an obvious hunting ground for these stellar populations. When looking at stars that are both metal poor and old, we again find that the majority of them ($\approx$ 60 per cent) can currently be found at radii greater than that of the solar neighbourhood ($>$ 8 kpc, see figure \ref{fig:XY}). However, if one could observationally disentangle all metal poor stars in the Milky Way from their more metal rich peers, the simulations suggest those at the centre are the most likely to be old and even to have formed before the epoch of reionization. Qualitatively, this result strengthens the conclusions of previous papers in this area that were based on a semi-analytic framework. Quantitatively though, many differences can be found that can be ascribed to the various prescriptions that are used. Most notably, we show in Fig.~\ref{fig:recipes} that quantitative results are sensitive to the metal mixing prescription.

The coming decade promises to be fruitful for the further study of the extremely metal poor and old Galaxy, with various surveys underway to mine these stars in various environments. One of the aims of this work is to provide guidance for any survey efforts designed to look for the most metal poor stars. We find that surveys of the outer halo are likely to uncover the sites of very early star formation in various environments. Roughly half of the stars formed before the epoch of reionization can be found at radii $> 8$ kpc today, as well as on average $\approx$ 75 per cent of the pristine stars (see Fig.~\ref{fig:XY4}). Whether these sites are truly free of the impact of the first stars forming elsewhere, either in the form of pollution with metals, or radiation or kinematic feedback, remains an open question that such surveys could help to constrain.

We find that for surveys focussing on the outer halo, it will be particularly rewarding to focus on stellar substructures -- either those that are still bound in the form of dwarf galaxies, or those that are remnants of stripping processes. In some of these substructures our model indicates that the number of truly old stars (formed within 0.8 Gyr after the Big Bang) among the metal poor populations could even be higher than in the Galactic centre (see Figures \ref{fig:foldpoorsatmains} and \ref{fig:XY4}). With regards to selecting dwarf galaxies for observational studies, our simulations suggest that the chance of studying purely old populations will be highest in the systems that were the most efficient star formers in the early Universe. Observationally these can be distinguished by their large fraction of old populations ($>10$~Gyr) and a \textit{low} fraction of metal poor stars. 

\bibliography{references}

\section*{Acknowledgements} 
We would like to thank Ryan Leaman, Nicolas Martin and Chris Brook for insightful discussions on this work, as well as the anonymous referee for thoughtful comments that helped to improve this manuscript. ES and KAO are grateful to the Mainz Institute for Theoretical Physics (MITP) for its hospitality and its partial support during the completion of this work. KAO is grateful for the generous hospitality of the Leibniz Institute for Astronomy Potsdam (AIP). ES gratefully acknowledges funding by the Emmy Noether program from the Deutsche Forschungsgemeinschaft (DFG). This work was supported by the Science and
Technology Facilities Council (grant number ST/F001166/1).
CSF acknowledges ERC Advanced Grant 267291 ``COSMIWAY''.
This work used the DiRAC Data Centric system at
Durham University, operated by the Institute for Computational
Cosmology on behalf of the STFC DiRAC HPC Facility
(www.dirac.ac.uk). This equipment was funded by BIS
National E-infrastructure capital grant ST/K00042X/1, STFC
capital grant ST/H008519/1, and STFC DiRAC Operations
grant ST/K003267/1 and Durham University. DiRAC is part
of the National E-Infrastructure. This research has made use of
NASA's Astrophysics Data System. The research was supported in part by the European Research
Council under the European Union's Seventh Framework
Programme (FP7/2007-2013) / ERC Grant agreement
278594-GasAroundGalaxies. RAC is a Royal Society University Research Fellow.

\appendix
\section{The effect of resolution}

\begin{figure}
\begin{center}
\includegraphics[width=\linewidth]{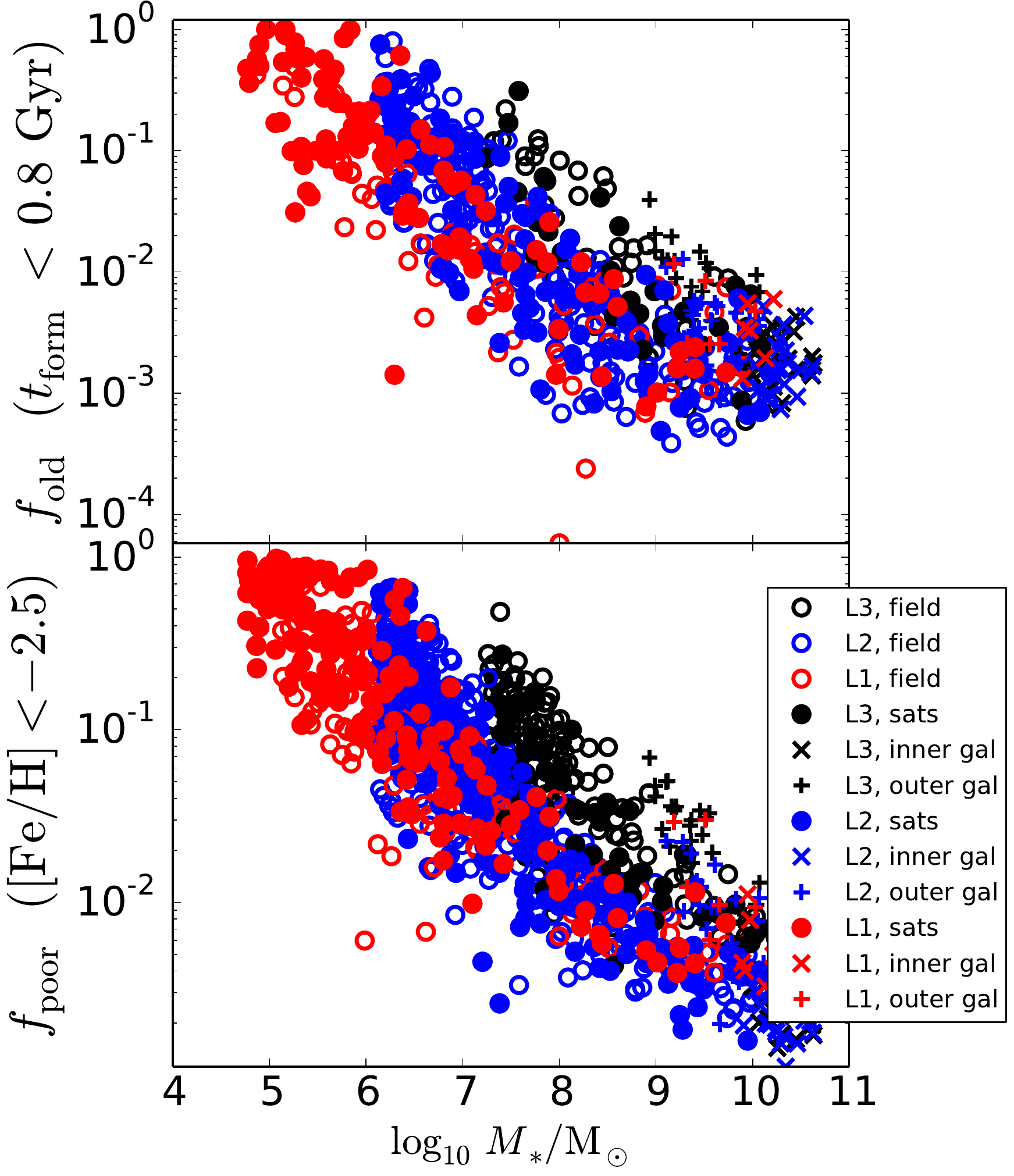}
\caption{The fraction of old (formation time $< 0.8$~Gyr after the Big Bang; top panel) or most metal poor ($[{\rm Fe}/{\rm H}] < -2.5$; bottom panel) populations as a function of stellar mass for galaxies within 2~Mpc of the main MW+M31-pair in each simulation. Filled symbols indicate galaxies within 300~kpc from either one of the main host galaxies. Black, blue and red points represent galaxies with the low- (L3), intermediate (L2), and high resolution (L1) simulations respectively.}
\label{fig:reseffects}
\end{center}
\end{figure}

Star formation processes and the general build-up of mass in galaxies may
depend on the simulation resolution. Three different
resolution versions have been run for the APOSTLE simulations, allowing us to
investigate the robustness of our results to variations in the particle mass. 

Fig.~\ref{fig:reseffects} shows the fraction of oldest or most metal poor populations (according to the definitions given in Section \ref{sec:oldormetpoor}) as a function of stellar mass for galaxies within 2~Mpc of the main MW/M31-pair in each simulation. Different colours are used for different resolution runs. From this figure it becomes clear that for the lowest-resolution simulations, resolution indeed plays a role. The galaxies generally have a higher fraction of oldest or respectively most metal poor stars than equal mass galaxies in higher-resolution runs. On the other hand, the intermediate and highest resolution simulations form a continuous sequence with large overlap. Additional analysis of the fraction of stars that are both metal poor and old -- a quantity that is central to our analysis in Section \ref{sec:oldandmetpoor} -- reveals that for all overlapping galaxy stellar mass bins the distribution of the fractions of oldest and most metal poor stars is consistent between the high and intermediate resolution simulations. A KS-test shows that the probability of the high- and intermediate-resolution distributions being drawn from a similar parent population is always larger than 0.18 for all mass bins.

\end{document}